\documentclass[12pt]{article}
\usepackage{amsfonts,amsmath}
\usepackage[hypertex]{hyperref}

\makeatletter \@addtoreset{equation}{section} \makeatother

\newcommand{\be}{\begin{equation}}
\newcommand{\ee}{\end{equation}}
\newcommand{\bee}{\begin{eqnarray}}
\newcommand{\beee}{\begin{array}}
\newcommand{\eee}{\end{eqnarray}}
\newcommand{\eeee}{\end{array}}

\newcommand{\ga}{\alpha}

\newcommand{\pb}{{\dot{\gb}}}

\newcommand{\gb}{\beta}
\newcommand{\gga}{\gamma}

\newcommand{\ie}{{\it i.e.,} }

\newcommand{\gd}{\delta}
\newcommand{\al}{\alpha}
\newcommand{\dal}{\dot{\alpha}}
\newcommand{\gl}{\lambda}
\newcommand{\gk}{\kappa}
\newcommand{\gep}{\epsilon}

\newcommand{\go}{\omega}

\newcommand{\q}{\,,\qquad}

\newcommand{\half}{\frac{1}{2}}

\newcommand{\p}{\partial}

\newcommand{\f}{\frac}

\newcommand{\PPP}{ {J} }

\newcommand{\dgb}{{\dot \gb}}
\newcommand{\dga}{{\dot \ga}}
\newcommand{\GGG}{\mathit{\Gamma}}

\newcommand{\dr}{{\rm d}}

\newcommand{\rmx}{{\mathrm{x}}}
\begin{document}
\begin{flushright}
FIAN/TD/10-17\\
\end{flushright}

\vspace{0.5cm}
\begin{center}
{\large\bf Test of the local form of higher-spin equations via
$AdS/CFT$}

\vspace{1 cm}

\textbf{V.E.~Didenko}\textbf{
and M.A.~Vasiliev}\\
 \vspace{0.5cm}
 \textit{I.E. Tamm Department of Theoretical Physics,
Lebedev Physical Institute,}\\
 \textit{ Leninsky prospect 53, 119991, Moscow, Russia}\\

\par\end{center}

\begin{center}
\vspace{0.6cm}
 didenko@lpi.ru, vasiliev@lpi.ru \\

\par\end{center}

\vspace{0.4cm}

\begin{abstract}
\noindent The local form of higher-spin equations found recently
to the second order \cite{Vasiliev:loc} is shown to
properly reproduce the anticipated $AdS/CFT$ correlators for
appropriate boundary conditions. It is  argued that consistent
$AdS/CFT$ holography for the parity-broken boundary models needs a
nontrivial modification of the bosonic  truncation of the original
higher-spin theory with the doubled number of fields, as well as a
nonlinear deformation of the boundary conditions in the higher
orders.
\end{abstract}

\section{Introduction}

Higher-spin (HS) theories (see e.g. \cite{Vasiliev:1999ba} for a
review)  have attracted much of interest providing a relatively
simple playground for $AdS/CFT$ correspondence
\cite{Maldacena:1997re}-\cite{Witten:1998qj}. Studying these
models may shed light on the nature of holography itself.
Particularly, some dualities relate complicated theory of gravity
and infinitely many HS fields in the bulk with simplest CFT duals
being just free theories. The HS $AdS/CFT$ story dates back to
stringy tensionless limit argument by Sundborg
\cite{Sundborg:2000wp} (see also \cite{WJ}-\cite{Sezgin:2002rt})
asserting free boundary theory as a HS dual.  A concrete proposal of Klebanov and Polyakov \cite{KP}, was
that what is known as HS $A$-model should be dual to either free
or critical $O(N)$-model. The conjecture was later generalized to
supersymmetric theories \cite{LP} and to HS $B$-model in
\cite{SS}. However, due to the lack of conventional action
principle for HS theory it was not clear how to test those
conjectures at the level of correlation functions until an
impressive calculation by Giombi and Yin \cite{GY1} based on a
certain setup for extracting tree-level correlators from equations
of motion. In \cite{GY1} and \cite{GY2} a substantial piece of
evidence in favor of the proposed dualities at the level of
three-point correlation functions was given. Later on Maldacena
and Zhiboedov showed \cite{MZh1} that the presence of infinitely
many exactly conserved HS currents in $d=3$ constrains CFT theory
drastically leaving one with either a theory of  free bosons or
free fermions. Yet, even if one allows for a slight HS symmetry
deformation, the CFT is still highly constrained \cite{MZh2}.
Though in this paper we  focus on the $AdS_4/CFT_3$ HS holography,
it should be noted that the important proposal on the
$AdS_3/CFT_2$ HS duality was put forward in
\cite{Gaberdiel:2010pz}.

Despite noticeable success of the $AdS_4/CFT_3$ HS holography tests
 some loose ends still remain even at
the level of three-point analysis especially in the sector of
holographic duality of parity-noninvariant $3d$ conformal theories
proposed in \cite{Aharony:2011jz,Giombi:2011kc} exhibiting
difficulties in extracting correlation functions from the
parity-noninvariant bulk HS theories \cite{GYrev} (where some were nevertheless
obtained). The main origin of those problems and
inconsistencies can be traced back to the nonlocal setup in HS
equations used in the original papers. Indeed, as was noticed in
\cite{GY1} the natural procedure of extracting HS interaction
vertices from HS equations results in a nonlocal interaction
even at the lowest nontrivial level leading to infinities in the
boundary limit. The origin of these nonlocalities is due to
natural ambiguity in field redefinitions in HS equations.
Particularly, the procedure of extracting HS vertices amounts to
solving some differential equations in the auxiliary spinorial
space which results in unavoidable problem of fixing a
representative. More generally, this is the problem of the choice of
proper (minimally nonlocal) class of functions respecting physical properties of nonlocal
theories such as HS theory.

Partly, the class of functions that respects nonlinear structure of
HS equations was proposed in \cite{Vasclass} and later further
narrowed in \cite{Vasiliev:loc,loc} for the special case of quadratic
corrections in the 0-form sector. In \cite{Vasiliev:loc} it was shown
that the proper field redefinition that brings HS equations into a
manifestly local form does exist, fixing relative coefficients of
the second order HS interaction vertices. In this paper we show
that the structure of second-order local HS interactions in four
dimensions is in perfect agreement with the CFT expectations.

The paper is organized as follows. In section \ref{HSeq} we
briefly review HS equations in four dimensions presenting
perturbative expansion up to the second order. Then, in section
\ref{Bndry} we discuss boundary conditions and  truncations
respecting the $AdS/CFT$ duality. In section \ref{corr} we extract
three-point correlation functions from the 0-form sector of HS
equations and in section \ref{conc} we leave our conclusion.

\section{Higher-spin equations}\label{HSeq}
HS equations in four dimensions have the  form \cite{more}
\begin{align}
&\dr W+W*W=0\,,\label{HS1}\\
&\dr S+[W,S]_*=0\,,\label{HS2}\\
&\dr B+[W,B]_*=0\,,\label{HS3}\\
&S*S=-i\theta_{\al}\wedge \theta^{\al}(1+F_*(B)*k*\gk)-
i\bar\theta_{\dal}\wedge \bar\theta^{\dal}(1+\bar F_*(B)*\bar k*\bar \gk)\,,\label{HS4}\\
&[S,B]_*=0\,.\label{HS5}
\end{align}
Here master fields $W(Z;Y;K|x)$, $B(Z;Y;K|x)$ and $S(Z;Y;K|x)$
depend on spinorial variables $Z_{A}=(z_{\ga}, \bar z_{\dga})$ and
$Y_A=(y_{\al}, \bar y_{\dga})$ ($\al, \dal=1,2$), as well as outer
Klein operators  $K=(k,\bar k)$. $W$ is a
space-time 1-form, $B$ is a 0-form and $S$ is a 1-form in the
exterior $Z^A$-directions with anticommuting differentials $\theta^A$. Functions of spinor variables
$Z_{A}=(z_{\ga}, \bar z_{\dga})$ and $Y_A=(y_{\al}, \bar
y_{\dga})$, $\al, \dal=1,2$ are treated as elements of the
star-product algebra with the associative star product

\be\label{star}
(f*g)(Z, Y)=\f{1}{(2\pi)^4}\int dU dV f(Z+U, Y+U)g(Z-V,
Y+V)e^{iU_A V^A}\,
\ee
($V^A=(\epsilon^{\ga\gb} V_\gb,\epsilon^{\dga\dgb} V_\dgb$).
Inner Klein operators
$\gk$ and $\bar\gk$ are
\be
\gk=e^{iz_{\ga}y^{\ga}}\,,\qquad \bar\gk=e^{i\bar z_{\dal}\bar
y^{\dal}}\,.
\ee
Outer Klein operator $k$ $(\bar k)$ is defined to anticommute with
all (anti)holomorphic variables
\be
\label{kout}
\{k,V_{\al}\}_*=0\,,\qquad k*k=1\,,
\ee
where $V_{\al}=(y_{\al}, z_{\al}, \theta_{\al})$. This formula extends
 the star product to $k,\bar k$-dependent elements.

 In this paper we focus on the  purely HS sector of the theory where
$B$ is linear in $k$ and $\bar k$ while $W$ and $S$ contain the $k,
\bar k$-independent part as well as bilinear $k*\bar k$.
Function $F_{*}(B)$ is set to be linear with an arbitrary constant complex
parameter
\be
F=\eta B\,,\qquad \bar F=\bar\eta B\,.
\ee
 Now, since
fields depend on outer Klein operators $k$ and $\bar k$, we
assume these to enter on the most right, for example,
\be
B(Z;Y; k, \bar k):=B(Z;Y)k+\bar{B}(Z;Y)\bar k\,.
\ee

System \eqref{HS1}-\eqref{HS5} can be analyzed perturbatively. One
starts with the vacuum solution that corresponds to pure $AdS_4$
space-time
\begin{align}
&B_0=0\,,\\
&S_0=Z_{A}\theta^A\,,\\
&W_0=\f i4(\go_{\al\al}y^{\al}y^{\al}+\bar\go_{\dal\dal}\bar
y^{\dal}\bar y^{\dal}+2e_{\al\dal}y^{\al}\bar{y}{}^{\dal})\,.
\end{align}
We take Poincare coordinates as well adopted for the boundary analysis
\be
\go_{\al\al}=-\f{i}{2z}d\rmx_{\al\al}\,,\quad
\bar{\go}_{\dal\dal}=\f{i}{2z}d\rmx_{\dal\dal}\,,\quad
e_{\al\dal}=\f{1}{2z}(d\rmx_{\al\dal}-i\gep_{\al\dal}dz)\,,
\ee
where $\rmx^{\ga\gb}=\rmx^{\gb\ga}$ denote the three boundary
coordinates (independently of whether they carry dotted or
undotted indices) while $z$ is the Poincar\'e coordinate.

First-order equations reduce to twisted-adjoint flatness condition
for the 0-form $B_1=C(Y; k, \bar k)$
\be\label{twad}
DC=D^{L}C+ie^{\al\dal}(y_{\al}\bar{y}_{\dal}-\p_{\al}\p_{\dal})C=0\,,
\ee
and First on-shell theorem for HS potentials $\go(Y)$
\be\label{oms}
D\go=\f i4\left(\eta \bar H^{\dal\dal}\p^{2}_{\dal}\bar C(0, \bar
y; k, \bar k)\bar k+\bar\eta H^{\al\al}\p_{\al}^2 C(y, 0; k, \bar
k)k\right)\,.
\ee
At second order the local form of HS equations was extracted from
\eqref{HS1}-\eqref{HS5} in \cite{Vasiliev:loc} for 0-form $C(Y)$
and in \cite{GelVas} for 1-form $\go(Y)$. The equation for 0-form
reads
\begin{align}\label{C2}
DC=\f i2\eta e^{\al\dal}\int e^{i\bar u_{\dal}\bar v^{\dal}}
y_{\al}(t\bar u_{\dal}+(1-t)\bar v_{\dal})J(ty, -(1-t)y, \bar
y+\bar u, \bar y+\bar v)k+c.c\,,
\end{align}
where
\be
J(y_1, y_2, \bar y_1, \bar y_2;k, \bar k):=C(y_1, \bar y_1; k,
\bar k)C(y_2, \bar y_2; k, \bar k)\,,
\ee
and we use the short-hand notation for integrals
\be
\label{int} \int F(t_1,\dots, t_n; \bar u, \bar
v):=\int\limits_{[0,1]^n}dt_1\dots dt_n\int\limits_{\mathbf{R}^4}
\f{1}{(2\pi)^2}d\bar u d\bar v F(t_1,\dots, t_n; \bar u, \bar
v)\,.
\ee
Similarly for integrals that contain both holomorhic $u, v$ and
antiholomorhic $\bar u, \bar v$ integration variables.

\section{Boundary conditions and truncations}\label{Bndry}
HS equations \eqref{HS1}-\eqref{HS5} admit various truncations.
Due to dependence on Klein operators $k$ and $\bar k$  there are
two copies of fields of every spin. In the bosonic case, the spectrum can be reduced
down to a single copy by setting
\be\label{bos}
B(Z,Y; k, \bar k)\to B(Z,Y)(k+\bar k)\,,\qquad W(Z,Y; k, \bar
k)\to W(Z,Y)(1+k\bar k)\,.
\ee
While bosonic truncation
\eqref{bos} can be imposed to all orders reducing the spectrum of
the theory, it  is not {\it a priori} guaranteed that it has any
CFT dual at all in the HS theories with broken parity
\cite{Vashol} . Within the perturbation theory however one can
impose condition relating fields of the full theory with the
doubled spectrum in such a way that the theory becomes bosonic yet
different from the one resulting from \eqref{bos}. To explain the origin
of the modified conditions driven by the $AdS/CFT$ requirement let
us analyze the boundary limit of the full fledged HS system in
perturbation theory.

\subsection{Lowest order}
Free-level analysis has been carried out in
\cite{Vashol}. According to it the field-current correspondence is
reached via the following identification
\be\label{T}
C(y, \bar y; k, \bar k)=z e^{y_{\ga}\bar y^{\ga}}T(w,\bar w; k,
\bar k)\,,
\ee
where
\be
w=\sqrt z y\,,\qquad \bar w=\sqrt z\bar y\,.
\ee
Eq. \eqref{T} says that if $C$ is on-shell, that is satisfies
\eqref{twad}, then $T$ enjoys the unfolded form of the $3d$ conformal current conservation equation
\be\label{tok}
\dr_{\rmx}T-\f i2\dr\rmx^{\ga\ga}\p_{\ga}\bar{\p}_{\ga}T=0\,.
\ee
HS potentials are sourced by the field  $C$ in the bulk in accordance
with \eqref{oms}. Its boundary pushforward reads
\be\label{1bndry}
D_{\rmx}\go_\rmx=\f14H^{\al\gb}_{\rmx\rmx}\f{\p^2}{\p w^{+\ga}\p
w^{+\gb}}\left(\bar\eta T(w^+,0)k-\eta T(0, iw^+)\bar k\right)\,.
\ee
One concludes that, in general, boundary HS fields, which
are gauge fields of the boundary conformal HS theory, are sourced by
currents. To make $AdS/CFT$ work in the standard sense, \ie for the
usual boundary CFT with the well-defined stress tensor,  one has to impose such boundary
conditions that make the right hand side of \eqref{1bndry} vanish
allowing to get rid of boundary HS gauge fields which can make the boundary stress
tensor gauge non-invariant.

For $\eta=1$ or $\eta=i$ proper conditions read
\be\label{AB}
T(w,\bar w)k=\pm T(-i\bar w, i w)\bar k\,.
\ee
It is important  that one can exclude scalar and spinor fields
from \eqref{AB} since they do not affect \eqref{1bndry} (at higher
orders this will not be the case) opening the way to alternative
boundary conditions in this sector, corresponding to the critical
boundary models in accordance with the original proposal of
\cite{KP}-\cite{SS}. Let us also note that there is no way to
include general parameter $\eta$ into \eqref{AB}  demanding
 $\eta=1$ or $\eta=i$.

However, for general $\eta$ conditions (\ref{AB})  can be modified
as follows. Having two fields in the decomposition
\be
C(Y; k, \bar k)=C(Y)k+\bar C(Y)\bar k
\ee
one can identify positive helicity component of a bosonic Weyl
module with $C(Y)$, while negative helicity part of the same field
with $\bar C(Y)$, i.e.
\be
C(Y):=C^{+}(Y)\,,\qquad \bar C(Y):=C^{-}(Y)\,,
\ee
where by the doubled helicity of a spin $s$ field we mean the difference
between the number of $y$ and $\bar y$ variables, in other words,
the eigenvalue of the following operator
\be\label{chir}
n=y^{\al}\f{\p}{\p y^{\al}}-\bar y^{\dal}\f{\p}{\p \bar
y^{\dal}}\,.
\ee
Particularly, $C^+$ carries more $y$ variables than $\bar y$ and
$C^-$ other way around. Let us stress that this way one truncates
the spectrum to the bosonic system in a way different from
\eqref{bos} allowing to get rid of the sources in \eqref{1bndry}
in the parity broken case by setting
\be\label{pbr}
\bar\eta T^{+}(w,\bar w)=\eta T^{-}(-i\bar w,i w)\,.
\ee

To make contact of the introduced boundary conditions with those
usually imposed in the HS literature consider HS boundary to bulk
propagators. In the 0-form sector the positive and negative helicity
parts have the following form \cite{GY1}
\be\label{prop1}
C^+=\eta
Ke^{if_{\al\dal}y^{\al}\bar{y}^{\dal}+i\xi^{\al}y_{\al}}\,,\quad
C^-=\bar\eta
Ke^{if_{\al\dal}y^{\al}\bar{y}^{\dal}+i\bar\xi^{\dal}\bar{y}_{\dal}}\,
\ee
(no $\eta$-factors for a scalar), where
\begin{align}
&K=\f{z}{(\rmx-\rmx_0)^2+z^2}\,,\\
&f_{\al\dal}=-\f{2z}{(\rmx-\rmx_0)^2+z^2}(\rmx-\rmx_0)_{\al\dal}-i\f{(\rmx-\rmx_0)^2-z^2}{(\rmx-\rmx_0)^2+z^2}\gep_{\al\dal}\,,\\
&\xi_{\al}=\Pi_{\al}{}^{\gb}\mu_{\gb}\,,\qquad
\Pi_{\al\gb}=K\left(\f{1}{\sqrt{z}}(\rmx-\rmx_0)_{\al\gb}-i\sqrt{z}\gep_{\al\gb}\right)\,,
\end{align}
and the reality conditions for polarization spinors are
\be
\mu_\al=i\bar\mu_{\al}\,.
\ee
While we will not use it in this paper, let us give for
completeness the explicit formula for 1-form $\go$
propagator\footnote{This form of the propagator was found by one
of us (V.D.) with Zhenya Skvortsov in 2014 but was never
published.}
\be\label{wprop}
\go=-\f i2Ke^{\al\dal}\xi_{\al}\bar\xi_{\dal}\int_{0}^{1}dt
e^{it\xi^{\al}y_{\al}+i(1-t)\bar\xi^{\dal}\bar y_{\dal}}\,.
\ee
Scalar part of the propagator \eqref{prop1} corresponds to the
$\Delta=1$ solution. Another scalar branch that stands for
$\Delta=2$ reads
\be\label{prop2}
C_{\Delta=2}=K^2(1+if_{\al\dal}y^{\al}\bar y^{\dal})\times
e^{if_{\al\dal}y^{\al}\bar y^{\dal}}\,.
\ee

Let us show how these propagators match different reality
conditions just spelled out. Using boundary prescription \eqref{T}
one finds for \eqref{prop1}
\begin{align}
&T^{+}=\f{\eta}{|\rmx-\rmx_0|^2}e^{-2i(\rmx-\rmx_0)^{-1}_{\al\al}w^{\al}\bar
w^{\al}+i(\rmx-\rmx_0)_{\al\gb}\mu^{\gb}w^{\al}}\,,\label{T+}\\
&T^{-}=\f{\bar\eta}{|\rmx-\rmx_0|^2}e^{-2i(\rmx-\rmx_0)^{-1}_{\al\al}w^{\al}\bar
w^{\al}+i(\rmx-\rmx_0)_{\al\gb}\bar\mu^{\gb}\bar
w^{\al}}\label{T-}
\end{align}
and
\be\label{sc2}
T_{\Delta=2}=\f{w_{\al}\bar w^{\al}}{|\rmx-\rmx_0|^4}\times
e^{-2i(\rmx-\rmx_0)^{-1}_{\al\al}w^{\al}\bar w^{\al}}\,.
\ee
One can see now that condition \eqref{AB} for $\eta=1$ is
fulfilled for \eqref{T+}, \eqref{T-} while for $\eta=i$ one has to
use \eqref{sc2} in accordance with parity-odd scalar condition for
HS $B$-model. For generic $\eta$ \eqref{T+} and \eqref{T-} as well
as \eqref{sc2} for alternative scalar boundary condition satisfy
\eqref{pbr}.

Having HS equations to the second order one may wish to examine
them in the boundary limit. Particularly, the expectation for
\eqref{AB} boundary condition for the $A$ and $B$ HS theories is
that in these cases HS symmetry
remains undeformed leading to conservation of boundary currents
yet leaving no HS gauge fields at the boundary. We will show that
this is indeed the case. For boundary conditions like \eqref{pbr}
or for alternative scalar like in the critical case on the
contrary it turns out that HS potentials get sourced on the
boundary and one should introduce certain nonlinear completion for
\eqref{pbr} at higher orders to make them vanish. This implies
among other things that without such a nonlinear completion  the
tree-level correlation functions extracted from the bulk are
anticipated to differ from boundary expectation starting from the
4-point functions.

\subsection{Second order}

Let us carry out boundary limit for \eqref{C2}. This  will give
 us the deformed version of current equation \eqref{tok}.
The limit is quite straightforward  using \eqref{T}. The
final result is
\be\label{cur}
\dr_{\rmx}T-\f {i}{2}\dr\rmx^{\ga\ga}\p_{\ga}\bar{\p}_{\ga}T=-\f
{\eta}{4}
\dr\rmx^{\ga\ga}w_{\ga}\int_{0}^{1}(t\bar{\p}_{2\ga}-(1-t)\bar\p_{1\ga})I\big(tw,
-(1-t)w, \bar w+i(1-t)w, \bar w-it w\big)k+c.c.\,,
\ee
where
\be
I(w_1, w_2, \bar w_1, \bar w_2)=T(w_1, \bar w_1; k, \bar k)T(w_2,
\bar w_2, k, \bar k)\,.
\ee
Note, that while field-current correspondence \eqref{T} contain
potentially dangerous projector $e^{iy_{\al}\bar y^{\al}}$ which
may cause infinities at the boundary it turns out that no
divergencies appear due to specific dependence on the homotopy
parameter $t$ in \eqref{C2}. One observes that currents receive
contributions originated from current-current interaction that may
lead to nonconservation. Indeed, from \eqref{cur} it follows that
\begin{align}\label{dT}
&\f{\p}{\p w^{\ga}}\f{\p}{\p w^{\gb}}\f{\p}{\p\rmx_{\ga\gb}}T=\notag\\
&=-\f {\eta}{4}\int_{0}^{1}dt\left(3+w^{\ga}\f{\p}{\p
w^{\ga}}\right)\left(t\bar\p^{\gb}_{2}-(1-t)\bar\p^{\gb}_{1}\right)\f{\p}{\p
w^{\gb}}\Big[T(tw, \bar w+i(1-t)w)T(-(1-t)w, \bar w-it
w)\Big]k+c.c.
\end{align}
which is nonzero in general resulting in
\be
\p\cdot J_s\neq 0\,.
\ee

Let us analyze this issue starting from the parity preserving models. In this case with the
boundary conditions \eqref{AB}  one finds that despite the deformation
is nonlinear the boundary currents remain conserved
\be\label{cons}
\f{\p}{\p w^{\ga}}\f{\p}{\p
w^{\gb}}\f{\p}{\p\rmx_{\ga\gb}}T=0\qquad \Rightarrow\qquad \p\cdot
J_s=0\,.
\ee
This can be most easily seen from noting that under \eqref{AB} the
right-hand side of \eqref{cur} gets rewritten as
\be\label{ABtok}
\dr_{\rmx}T-\f i2\dr\rmx^{\ga\ga}\p_{\ga}\bar{\p}_{\ga}T=-\f
{\eta}{4} \dr\rmx^{\ga\ga}w_{\ga}\f{\p}{\p w^{\ga}}\int_{0}^{1} dt
I\big(tw, -(1-t)w, \bar w+i(1-t)w, \bar w-it w\big)\,,
\ee
from where \eqref{cons} immediately follows. The fact that for
free theories \eqref{ABtok} results in current conservation means
that there is a local field redefinition that brings \eqref{ABtok}
to the canonical conserved current form \eqref{tok}. For parity
broken boundary condition \eqref{pbr} the HS currents no longer
conserve. In obtaining \eqref{ABtok} the structure of \eqref{cur}
was important. Particularly one uses the symmetry with respect to
the exchange $t\to 1-t$. The check carried out for parity
preserving boundary conditions \eqref{cons} alone is sufficient to
justify the agreement between bulk vertices given in \eqref{C2}
and boundary free theory 3pt correlation functions. Indeed,
according to Maldacena-Zhiboedov theorem \cite{MZh1} the
conservation of HS currents inevitably implies free boundary
theory.

A soft spot in this argument is the following. As a matter of
principal it may happen that while HS currents do conserve the
theory still contains sources for the boundary HS connections, in
which case the standard $AdS/CFT$ correspondence can be lost.
 So let us check out the conditions at which sources for HS
connections do vanish. To do so we should analyze the 1-form
sector found in \cite{GelVas} in the boundary limit.

It is easy to perform boundary limit for current interaction
equation in the 1-form sector of \cite{GelVas} following the logic of
\cite{Vashol}, arriving at the equation
\begin{align}\label{1-bndr}
&D_{\rmx}\go_{\rmx}(w^+, v^-)=\f i8\eta\bar\eta\int d^2
t\gd'(1-t_1-t_2)H_{\rmx\rmx}^{\ga\ga}\left(\f{\p}{\p
u^{\ga}}\right)^2\times\\
&\times \Big\{I(t_1(w^++u), -t_2(w^++u), it_2w^+, -it_1w^+)
-I(t_1w^+, -t_2w^+, it_2(w^++u),
-it_1(w^++u))\Big\}\Big|_{u=0}\,,\notag
\end{align}
where the following variables have been introduced
\be
w=w^{+}+izv^-\,,\qquad \bar w=iw^++zv^-\,.
\ee
Just as well as at the linearized  level, one observes that sources for HS
connections do not vanish in general (although they almost do
since the two terms on the right-hand side of \eqref{1-bndr} are equal
to each other at $u=0$). However, imposing free theory boundary
conditions \eqref{AB} one finds exact cancellation and the theory
becomes free of boundary HS connections in accordance with the
$AdS/CFT$ expectation.

An important comment is now in order. One may expect that for
parity breaking boundary conditions \eqref{pbr} or for those of
critical theories one has vanishing sources for boundary
connections too. This is not the case as \eqref{AB} is likely to be
the only linear relation that cancels out sources. This implies
that on the way of proposed dualities with critical models and
vectorial models with Chern-Simons matter one has to modify
boundary conditions to compensate the nonlinear corrections.

Also it  should be noted that in the CFT-based HS literature terms on
the r.h.s. of (\ref{cur}) are usually interpreted as ``slight
breaking" of HS symmetry \cite{MZh2}. From the perspective of the
original HS equations, however, they are naturally interpreted as
a deformation rather than breaking of HS symmetry. Indeed,
consistency  of nonlinear terms on the r.h.s. of HS field
equations implies that  the HS gauge symmetry transformations
receive nonlinear corrections as well. The tricky point is that,
at the boundary,  the resulting deformation may go beyond the
standard class of CFTs with well defined (gauge invariant) stress
tensor because the deformed HS gauge transformation in most cases
mixes HS 0-forms, that have clear meaning from the boundary CFT
perspective, with the HS 1-forms at the boundary, which are
conformal HS gauge fields on the boundary not allowed in the
standard CFTs.

\section{Boundary correlators }\label{corr}
In this section we venture to extract correlation functions of a
dual theory from the 0-form sector of the bulk field equations. We
restrict ourselves to the case with two sources on the boundary
$s_1$ and $s_2$ that generate spin $s$ such that
\be\label{ss}
s\geq s_1+s_2\,.
\ee
This constraint comes from the fact that so far we have taken
into account only current interactions given in \eqref{C2}, which
is only consistent when restriction \eqref{ss} is imposed since otherwise
the contribution of HS 1-forms also has to be taken into account. For the opposite
case of three spins obeying the triangle inequalities, the original current
interaction is supported by the HS 1-forms and is local in the original setup of
 Giombi and Yin \cite{GY1}, giving  the proper  answer.

According to the proposal of  \cite{GY1}
a solution to the second order equation for Weyl 0-form
\eqref{C2} generated by two boundary sources can be associated
with a properly normalized 3pt function via
\be
\langle JJJ\rangle\sim\lim_{z\to 0}z^{-1}G(wz^{-\f 12}, \bar
wz^{-\f 12})\Big|_{\bar w=0}\,,
\ee
where $G(y,\bar y)$ is a Green's function for equation \eqref{C2}.
The remaining $w$-variable is to be associated with a
polarization spinor for the outgoing leg of spin $s$. Though such a
prescription for the computation of correlation functions may need some
further justification, for a time being we take it as a working tool.
Before going into technical details of the computation we give  general
arguments on the dependence of the boundary correlators on the phase parameter
in the HS theory.

\subsection{Phase dependence}
\label{holo} In this section we reconsider the analysis of HS
holography of \cite{Vasiliev:loc} in a more conventional setup
leading  to the same conclusions. To this end, consider HS
equations of \cite{GelVas}
\be \label{W2hhloc0sch}
D\omega (y,\bar y) = \f{ i}{4} \Big ( \eta
\bar{H}^{\dga\pb}\f{\p^2}{\p \overline{y}^{\dga} \p
\overline{y}^{\dgb}}\ {C  }_-(0,\overline{y}| x) +\bar \eta
H^{\ga\gb} \f{\p^2}{\p {y}^{\ga} \p {y}^{\gb}}\ {C }_+(y,0| x)\Big
) + \eta\bar\eta  {\GGG}^{loc}(\PPP)\,,\,\qquad
\ee
where $\GGG^{loc}$ is the second-order current interaction,
$C_\pm$ denote positive and negative helicity parts of $C(y,\bar
y)$ and the dependence on the Klein operators is discarded. Though
as shown in \cite{Vasiliev:loc,GelVas} the quadratic
$\PPP$-dependent deformation is independent of the phase of
$\eta=|\eta|\exp{i\varphi}$,
the linear part is phase-dependent. Introducing the new fields
\be
C'_- = \eta C_-\q C'_+ = \bar \eta C_+\,,
\ee
Eq. \eqref{W2hhloc0sch} takes the form
\be \label{W2hhloc0'}
D\omega(y,\bar y) = \f{ i}{4} \Big (  \bar{H}^{\dga\pb}\f{\p^2}{\p
\overline{y}^{\dga} \p \overline{y}^{\dgb}}\
{C}_-'(0,\overline{y}| x) + H^{\ga\gb} \f{\p^2}{\p {y}^{\ga} \p
{y}^{\gb}}\ {C }_+'(y,0| x)\Big ) +
{\GGG}^{loc}(\PPP(C(C')))\,,\,\qquad
\ee
where, setting for simplicity $|\eta|=1$,
\be
\label{em}
C(C') = \exp i\varphi \,C_+' + \exp -i\varphi\, C_-'\,.
\ee
Clearly,  redefinition (\ref{em}) is an $U(1)$ electromagnetic duality
transformation with the phase $\varphi$.

The linear term in Eq.~(\ref{W2hhloc0'}) tells us that it is the 0-form $C'$
that has to be identified with the generalized Weyl (Faraday for $s=1$) tensor
associated with the curvatures of the Fronsdal HS fields contained in $\go(y,\bar y)$.
In these terms, the vertex which was $\varphi$-independent in terms of $C$ acquires
the nontrivial $\varphi$-dependence in terms of $C'$
\be\label{G}
{\GGG}^{loc}(\PPP) =  {\GGG}^{loc} \left(\exp 2i\varphi J_{++}(C')
+
 \exp -2i\varphi J_{--}(C')
 + 2 J_{+-}(C')\right)\,.
\ee
Since the $A$-model with $\varphi=0$ and $B$-model with $\varphi=\frac{\pi}{2}$ are known to
correspond to  bosonic and fermionic parity-invariant boundary vertices,  we  set
\be\label{bf}
J_b := J_{++}(C') +
  J_{--}(C')
 + 2 J_{+-}(C')\q J_f := - J_{++}(C') -
  J_{--}(C')
 + 2 J_{+-}(C')\,.
\ee
The remaining parity-odd boundary vertex is associated with
\be\label{podd}
J_o = i (J_{++}(C') - J_{--}(C'))\,.
\ee
 In terms of these currents, ${\GGG}^{loc}(\PPP)$  acquires the form
\be\label{phdep}
{\GGG}^{loc}(\PPP) = \cos^2 (\varphi ) J_b + \sin^2 (\varphi ) J_f
+\half \sin(2\varphi) J_o\,
\ee
coinciding with the expression obtained in \cite{Vasiliev:loc} by
slightly different arguments. This expression precisely matches
the dependence on the phase $\varphi$ anticipated from the HS
holography \cite{MZh2,Aharony:2011jz,Giombi:2011kc,GYrev}.

 To
summarize, the proper phase dependence of the current interactions
in the phase-independent vertex results from that in the terms
linear in the 0-forms upon the identification of the genuine HS
Weyl tensors.

This simple analysis is useful in many respects. In particular it shows that, to find
the phase dependence of the boundary correlators it suffices to know it for any three
different data in $\varphi$. For instance it is enough to find the boundary correlators
in the $A$-model with $\varphi=0$, $B$-model with $\varphi=\pi/2$ to identify the
parity-even part, and, say, the first $\varphi$-derivative at its $B$-model value
$\varphi=\pi/2$ to identify its parity-odd part. Interestingly, the latter definition
is somehow reminiscent of the interpretation of the odd $3d$ conformal structure proposed in
\cite{Bonora:2016ida} as a massive deformation of $3d$ fermionic currents. Indeed,
from the boundary perspective the parameter $\eta$ is closely related to the $3d$ massive
boundary deformation though at a nonzero VEV of the 0-form B, which, though making
sense in the model including topological fields not considered in this paper,
has to be set to zero in the end of the computation.

\subsection{0-form Green's function}
The righthand side of equation \eqref{C2} contains two pieces
proportional to $\eta$ and  $\bar\eta$, respectively. Therefore
the Green's function can be found as a sum of two
\be
G=G_\eta+G_{\bar\eta}\,,
\ee
where $G_\eta$ (similarly $G_{\bar\eta}$) obey the equation
\begin{align}
DG_{\eta}=\f i2\eta e^{\al\dal}\int e^{i\bar u_{\dal}\bar
v^{\dal}} y_{\al}(t\bar u_{\dal}+(1-t)\bar v_{\dal})J(ty, -(1-t)y,
\bar y+\bar u, \bar y+\bar v)k\,.
\end{align}

In terms of power series, the Green's function was analyzed in
\cite{loc}. Here we would like to have its representation
suitable for practical calculations. So, let us use the following
Ansatz for $G_\eta$,
\be\label{Gansatz}
G_\eta=\eta\int  f(t_1,t_2,t_3)e^{iu_Av^A}J(u+t_1y, t_3v-t_2y,
\bar y+\bar u, \bar y+\bar v)k\,,
\ee
which is most general in the  holomorphic sector of spinor variables.
Since the measure in (\ref{int}) is compact and assuming that a function (distribution)
$f(t_1, t_2, t_3)$ is well defined we will be freely integrating by parts.
Substituting \eqref{Gansatz}
into \eqref{C2} one finds (for more detail see \cite{loc})
\be
f(t_1, t_2, t_3)=\f12\gd'(1-t_1-t_2-t_3)\,,
\ee
and, therefore,
\be\label{G}
G_\eta=\f{\eta}{2}\int\gd'(1-t_1-t_2-t_3)e^{iu_Av^A}J(u+t_1y,
t_3v-t_2y, \bar y+\bar u, \bar y+\bar v)k\,.
\ee
So defined Green's function does not satisfy \eqref{C2} in general
in the first place  because \eqref{C2} is not everywhere
consistent in particular because the contribution of 1-forms
should be taken into account if the constraint \eqref{ss} is
not respected. But even for $s\geq
s_1+s_2$ when \eqref{ss} is fulfilled the Green's function,
\eqref{G} is only valid for those sources in which constituent fields
$C_1(y,\bar y)$ and $C_2(y, \bar y)$  have opposite chiralities \eqref{chir}, \ie
\be
n(C_1) n(C_2)<0\,.
\ee
Since the extension of the Green's function to the general case
with arbitrary signs of chiralities yet remains to be constructed,
our strategy will be as follows. Assuming that the coefficients in
correlation function depend solely on spins in the vertex, \ie modules of
helicities, we will take sources (\ie constituent fields) of opposite
chiralities in the calculation.

There are two sets of primary currents stored in the boundary
limit of the Green's function: those depending only on $w$
or only on $\bar w$. We focus on the $w$-dependent ones   which
makes it possible extracting correlation functions from the
holomorphic part $G_{\eta}$. In accordance with the general analysis
of section \ref{holo}, there are three different structures that
arise upon substituting propagators \eqref{prop1} that satisfy
boundary conditions \eqref{pbr}
\begin{align}
&\langle JJJ\rangle_{boson}\sim G^{++}+G^{--}+G^{+-}+G^{-+}\,,\label{c1}\\
&\langle JJJ\rangle_{fermion}\sim G^{++}+G^{--}-G^{+-}-G^{-+}\,,\label{c2}\\
&\langle JJJ\rangle_{odd}\sim G^{++}-G^{--}\,,\label{c3}
\end{align}
where pluses and minuses denote chirality signs. The dependence on
the phase parameter $\eta$ is fixed according to \eqref{phdep}.
Particularly, it follows that $G^{++}+G^{--}$ and $G^{+-}+G^{-+}$
correspond to free theories correlators. Substituting propagators
\eqref{prop1} into \eqref{G} and performing simple Gaussian
integration leads to the following result in the leading order in
$z$
\begin{align}
&G^{+-}_{12}=\int d^3t\f{K_1
K_2}{\Delta}\gd'(1-t_1-t_2-t_3)e^{2\f{t_{1}t_{2}}{\Delta}Q+\f{t_1}{\Delta}((1-t_3)P_1+zt_3\tilde{S}_1)
-\f{t_1}{\Delta}((1-t_3)P_2+zt_3\tilde{S}_2)}\,,\label{+-}\\
&G^{-+}_{12}=\int d^3t\f{K_1
K_2}{\Delta}\gd'(1-t_1-t_2-t_3)e^{2\f{t_{1}t_{2}}{\Delta}Q+\f{t_2}{\Delta}((1-t_3)P_1-zt_3\tilde{S}_1)
-\f{t_2}{\Delta}((1-t_3)P_2-zt_3\tilde{S}_2)}\label{-+}\,.
\end{align}
Here, indices 1 and 2 label points  at the boundary and
\be
\Delta=(1-t_3)^2+z^2\gep^2 t_3+O(z^4)\,,\qquad
\gep=\f{2\,\rmx_{12}}{\rmx_{01}\rmx_{02}}\,,
\ee
where the outgoing leg $\rmx$ is denoted by $\rmx_0$
\be
\rmx:=\rmx_0\,.
\ee
The parity-preserving conformal structures are denoted by $P$ and
$Q$, while $\tilde S$ denote parity-odd ones. We specify these later
upon taking the boundary limit.

An important comment is that, naively, it looks like in the
boundary limit $z\to 0$ all parity-odd structures $\tilde S$
vanish, because they are accompanied by a factor of $z$. This is
not the case due to the pole at $z=0$ resulting from $\Delta$ upon
integration over $t_3$. Indeed,  careful analysis shows that the terms
$(1-t_3)$ and $zt_3$ in exponentials \eqref{+-} and \eqref{-+} give the
same contribution in $z$ as $z\to 0$. Carrying out the boundary
limit $z\to 0$ and integrating over $t_1$ and $t_2$ one arrives at
the following result
\begin{align}
G^{++}_{12}=\f z2
K_{s_1s_2s}\f{Q^{s-s_1-s_2}}{|\rmx_{01}||\rmx_{02}||\rmx_{12}|}\int_{0}^{\infty}
d\tau\f{\tau^{2s}(\tau P_1+S_1)^{2s_1}(-\tau
P_2+S_2)^{2s_2}}{(1+\tau^2)^{s+s_1+s_2+1}}\,,\label{G++}\\
G^{--}_{12}=\f z2
K_{s_1s_2s}\f{Q^{s-s_1-s_2}}{|\rmx_{01}||\rmx_{02}||\rmx_{12}|}\int_{0}^{\infty}
d\tau\f{\tau^{2s}(\tau P_1-S_1)^{2s_1}(-\tau
P_2-S_2)^{2s_2}}{(1+\tau^2)^{s+s_1+s_2+1}}\,,\\
G^{+-}_{12}=\f z2
K_{s_1s_2s}\f{Q^{s-s_1-s_2}}{|\rmx_{01}||\rmx_{02}||\rmx_{12}|}\int_{0}^{\infty}
d\tau\f{\tau^{2s}(\tau P_1+S_1)^{2s_1}(-\tau
P_2-S_2)^{2s_2}}{(1+\tau^2)^{s+s_1+s_2+1}}\,,\\
G^{-+}_{12}=\f z2
K_{s_1s_2s}\f{Q^{s-s_1-s_2}}{|\rmx_{01}||\rmx_{02}||\rmx_{12}|}\int_{0}^{\infty}
d\tau\f{\tau^{2s}(\tau P_1-S_1)^{2s_1}(-\tau
P_2+S_2)^{2s_2}}{(1+\tau^2)^{s+s_1+s_2+1}}\,,\label{G-+}\\
\end{align}
where
\begin{align}
K_{s_1s_2s}=\f{2^{s-s_1-s_2}(s+s_1+s_2)!}{(2s)!(2s_1)!(2s_2)!}\,.
\end{align}
Note that for equal chirality signs, i.e. for $G^{++}$ and
$G^{--}$, the coefficient $K_{s_1s_2s}$ would be different should
we still used \eqref{G} in this case as a Green's function.

The conformal structures appear in the following combinations
\begin{align}
&P_1=i\f{(\rmx_{01})_{\ga\ga}w^{\ga}\mu_{1}^{\ga}}{|\rmx_{01}|^{2}}\,,\qquad
P_2=i\f{(\rmx_{02})_{\ga\ga}w^{\ga}\mu_{2}^{\ga}}{|\rmx_{02}|^{2}}\,;\qquad
Q=\left(\f{\rmx_{01}}{|\rmx_{01}|^2}-\f{\rmx_{02}}{|\rmx_{02}|^2}\right)_{\ga\ga}w^{\ga}w^{\ga}\,,\label{conf1}\\
&S_{1}=\f{(\rmx_{02})^{\gb\al}(\rmx_{12})_{\al}{}^{\gga}\mu_{1\gga}w_{\gb}}{|\rmx_{01}||\rmx_{02}||\rmx_{12}|}\,,\qquad
S_{2}=\f{(\rmx_{01})^{\gb\al}(\rmx_{12})_{\al}{}^{\gga}\mu_{2\gga}w_{\gb}}{|\rmx_{01}||\rmx_{02}||\rmx_{12}|}\,.\label{conf3}\\
\end{align}
To identify three-point correlation functions from
\eqref{G++}-\eqref{G-+} one uses prescription
\eqref{c1}-\eqref{c3} and symmetrization over the sources at
$\rmx_1$ and $\rmx_2$.

As noted above, $G^{++}+G^{--}$ and $G^{+-}+G^{--}$ correspond to
the parity-preserving three-point functions. To verify these
against free theory correlators let us start with
\be\label{b+f}
G^{+-}_{12}+G^{-+}_{12}= \f z2
K_{s_1s_2s}\f{Q^{s-s_1-s_2}}{|\rmx_{01}||\rmx_{02}||\rmx_{12}|}\int_{-\infty}^{\infty}
d\tau\f{\tau^{2s}(\tau P_1+S_1)^{2s_1}(\tau
P_2+S_2)^{2s_2}}{(1+\tau^2)^{s+s_1+s_2+1}}\,.
\ee

Naively it may seem that \eqref{b+f} has nothing to do with
correlators of currents of free boson and free fermion as it
depends on the parity-odd structure $S$. However, since the
integration in \eqref{b+f} is carried out along the real axis, the
parity-odd structures will appear in bilinear combinations leading
to a
 parity-even result. Since conformal structures
\eqref{conf1}-\eqref{conf3} are not algebraically independent
(see e.g., \cite{GY3} for a list of identities on
these structures), it is hard to identify
in this expression the product of cosines and sines found in
\cite{GY2}.
(The form of the final result is sensitive to a particular
representation choice.) For a simple check showing that the result
matches free theory correlators it is convenient to fix boundary
points as follows
\be
\rmx_0=0\,,\quad \rmx_1=\rmx\,,\quad \rmx_2=\rmx-\delta\,,\quad
|\delta|\ll |\rmx|
\ee
and take equal polarization spinors
\be
w_{\ga}=i\mu_{1\ga}=i\mu_{2\ga}=\gl_{\ga}\,.
\ee
In addition it is convenient to require
\be
\overrightarrow{x}\cdot\overrightarrow{\gl}=0\,.
\ee
In this limit, which was also used in \cite{GY1} and is similar to the
light cone limit of \cite{MZh1}, the 3pt correlators calculated in
a free theory amount to \cite{GY1}
\be\label{O(N)}
\langle J_{s_1}(\rmx, \gl)J_{s_2}(\rmx-\gd, \gl)J_{s}(0,
\gl)\rangle_{\gl\cdot\rmx=0}\sim\f{\Gamma(s_1+s_2+\f12)\Gamma(s+\f12)}{\pi
s_1!s_2!s!}\f{(\gl\cdot\gd)^{s_1+s_2+s}}{|\rmx|^{2s+2}|\gd|^{2s_1+2s_2+1}}\,.
\ee

Let us see what \eqref{b+f} gives in this limit. From
\eqref{conf1}-\eqref{conf3} one finds
\begin{align}
&P_1=0\,,\qquad P_{2}=\f{\gd\cdot\gl}{|\rmx|^2}\,,\qquad
Q=-\f{\gd\cdot\gl}{|\rmx|^2}\,,\\
&S_1=S_2=i\f{(\rmx\cdot\gd)_{\ga\ga}\gl^{\ga}\gl^{\ga}}{|\rmx|^2|\gd|}\,.
\end{align}
Since $\gd\ll\rmx$ we can neglect $P_2$ in \eqref{b+f} and
therefore
\be
G^{+-}_{12}+G^{-+}_{12}= \f z2
K_{s_1s_2s}\f{Q^{s-s_1-s_2}S_1^{2s_1+2s_2}}{|\rmx|^2|\gd|}\int_{-\infty}^{\infty}
d\tau\f{\tau^{2s}}{(1+\tau^2)^{s+s_1+s_2+1}}\,.
\ee
Using Fierz (\ie Schoutens) identities it is easy to see, that
(note, that \eqref{G++}-\eqref{G-+} do not apply for half-integer
spins)
\be
S_1^2=-\f{(\gd\cdot\gl)^2}{|\rmx|^2|\gd|^2}
\ee
leading to
\be
G^{+-}_{12}+G^{-+}_{12}\sim
K_{s_1s_2s}\f{(\gl\cdot\gd)^{s_1+s_2+s}}{|\rmx|^{2s+2}|\gd|^{2s_1+2s_2+1}}
\int_{-\infty}^{\infty}
d\tau\f{\tau^{2s}}{(1+\tau^2)^{s+s_1+s_2+1}}\,.
\ee
Integrating by residues,
\be
\int_{-\infty}^{\infty}
d\tau\f{\tau^{2s}}{(1+\tau^2)^{s+s_1+s_2+1}}=\f{\Gamma(s+\f12)\Gamma(s_1+s_2+\f12)}{(s+s_1+s_2)!}
\ee
and substituting $K_{s_1s_2s}$ one finds
\be
G^{+-}_{12}+G^{-+}_{12}\sim
\f{\Gamma(s+\f12)\Gamma(s_1+s_2+\f12)}{(2s)!(2s_1)!(2s_2)!}\f{(\gl\cdot\gd)^{s_1+s_2+s}}{|\rmx|^{2s+2}|\gd|^{2s_1+2s_2+1}}\,,
\ee
which is consistent  with the free theory
prediction \eqref{O(N)} upon an appropriate 2pt-normalization. Same is true for
$G^{++}_{12}+G^{--}_{12}$.

The parity-odd contribution resides in
$G_{12}^{++}-G_{12}^{--}$ and the corresponding three-point function can be
obtained from that expression by symmetrizing sources at $\rmx_1$
and $\rmx_2$. Up to the two-point function normalization $\langle
J_sJ_s\rangle$ the final result reads
\begin{align}\label{odd}
&{\langle
J_{s_1}J_{s_2}J_{s}\rangle_{odd}}\sim\f12\f{K_{s_1s_2s}}{|\rmx_{01}||\rmx_{02}||\rmx_{12}|}\int_{0}^{\infty}
d\tau\f{\tau^{2s}}{(1+\tau^2)^{s+s_1+s_2+1}}Q^{s-s_1-s_2}\times\\
&\left( (\tau P_1+S_1)^{2s_1}(-\tau P_2+S_2)^{2s_2} -(\tau
P_1-S_1)^{2s_1}(-\tau P_2-S_2)^{2s_2}\right)+(\rmx_1, \mu_1,
s_1)\leftrightarrow(\rmx_2, \mu_2, s_2)\,.\notag
\end{align}
Recall that spins are restricted by (\ref{ss}). To see that the result is
nonzero it is enough to consider the case of $s_1=1$, $s_2=0$ which gives
\be
\langle O_{\Delta=1}(\rmx_2)J_1(\rmx_1)J_{s}(\rmx_0)\rangle\sim
\f{s!}{(2s)!}\f{2^{s-2}}{|\rmx_{01}||\rmx_{02}||\rmx_{12}|}
(Q^{s-1}+(-Q)^{s-1})P_1S_1\,.
\ee
Similarly, using \eqref{sc2} propagator one can calculate
correlation functions corresponding to critical models. We do not
perform this calculation in our paper. Note that from the boundary
side nonconservation of HS currents in the parity-broken case was
recently studied in \cite{Giombi:2016zwa}, where some correlators
were explicitly found. It will be interesting to compare them with
\eqref{odd}.

As stressed earlier, the form of the final result \eqref{odd}
heavily depends on the freedom in using  relations on conformal
structures \eqref{conf1}-\eqref{conf3}. We expect \eqref{odd} to
admit a simpler representation. In this respect it is interesting
to note that typical integrals that show up in the boundary limit
of a Green's function
\be\label{int}
G(C_{s_1}, C_{s_2})\sim
\int_{0}^{\infty}d\tau\f{\tau^{2s}}{(1+\tau^2)^{s+s_1+s_2+1}}(\tau
a+b)^{2s_1}(\tau c+d )^{2s_2}\,,
\ee
where $a, b, c, d$ are some conformal structures among list
\eqref{conf1}-\eqref{conf3}, can be rewritten upon the change of integration
variable  $\tau=\tan\phi$ as
\be
R\int_{0}^{\pi/2}d\phi\,\sin^{2s}\phi\,\sin^{2s_1}(\phi+\phi_1)\sin^{2s_2}(\phi+\phi_2)\,,
\ee
where
\be
R=(a^2+b^2)^{s_1}(c^2+d^2)^{s_2}\,,\qquad \tan\phi_1=\f
ba\,,\qquad \tan\phi_2=\f dc\,.
\ee
This representation may be useful for finding  a simpler
representation for the parity-odd three-point functions.

\section{Conclusion}\label{conc}
The main findings of our work are the following. We have examined
local form of HS equations to the second order at the level of
equations of motion in its most sensitive part of the current
interaction sector with spins obeying $s\geq
s_1+s_2$, \ie outside the triangle inequality region. We have
checked whether the coefficients obtained in \cite{Vasiliev:loc}
and \cite{GelVas} are consistent with the boundary theory expectations
and found perfect agreement. Particularly, the boundary limit that
describes deformation to current conservation condition is
consistent with the requirement for free theories to have exactly
conserved HS currents. For these boundary conditions we have also
checked that, in agreement with the conventional $AdS/CFT$ prescription,
 no HS connections survive at the boundary. These facts,
 being crucially
dependent on the structure of vertices obtained in
\cite{Vasiliev:loc}, confirm that the prescription of
\cite{Vasiliev:loc} is the only proper one. Still we have carried
out some calculation at the level of three point functions
extracted from the 0-form sector {\it a la} Giombi and Yin \cite{GY1}.
Though details of the prescription of extracting correlators from the 0-form
sector of HS equations is not entirely clear to us and
perhaps needs some further analysis (particularly, this concerns the argument on the
linear relation between the Weyl module and HS connections) we found perfect agreement in
case of free theories. For parity broken case we have calculated
correlation functions $\langle J_{s_1}J_{s_2}J_{s_3}\rangle_{odd}$
for $s_{3}\geq s_1+s_2$ using the same approach. The result is
nonzero which seemingly contradicts to the analysis of \cite{MZh1}
where parity-odd three-point functions were found within the
triangle identity $s_i\leq s_j+s_k$ and it was claimed that for
$s_3\geq s_1+s_2$ the result is zero. However, it is important to
note, that in paper \cite{MZh1} all HS currents were supposed to
be conserved, while in our case we do not have current
conservation for parity-odd case which is in agreement with
general analysis of \cite{GY3}.

Another observation of our work highlighting the conjecture of
\cite{Vashol} on the role of the boundary conditions is that,
apart from the case of free boundary theories, no boundary
conditions linear in the HS 0-forms make the sources to the
boundary HS connection vanish. Particularly, for critical models
and parity-broken models a nonlinear correction to the source for
boundary connections always springs out. This implies that
boundary conditions consistent with the standard $AdS/CFT$
prescription may need a nonlinear deformation anticipated to
become important starting from the 4pt correlation functions. This
deformation is similar to the one observed recently in
$\mathcal{N}=8$ supergravity theory which requires boundary
supersymmetry modification in order to match superconformal
correlation functions \cite{Freedman:2016yue}.

Finally, the analysis carried out in this paper suggests that even in the purely bosonic case
there exist perturbatively different reductions of the full
nonlinear HS equations with the doubled set of fields compared to the naive
reduction with a single set of bosonic fields of any spin.

\section*{Note added} After completion of our work we learned that
closely related problem was considered in \cite{SezSkv} by E.
Sezgin, E.D. Skvortsov and Y. Zhu

\section*{Acknowledgements}
We acknowledge  a partial support from  the Russian Basic
Research Foundation Grant No 17-02-00546. MV is grateful to
 the Galileo Galilei Institute for Theoretical Physics (GGI) for the hospitality and INFN for
 partial support during the completion of this work within the program "New Developments in AdS3/CFT2 Holography".
 The work of MV is partially supported by a grant from the Simons Foundation.
V.D. is grateful to professor Hermann Nicolai for kind hospitality
at AEI where a part of this work has been done. We are grateful to
Olga Gelfond for important comments, Simone Giombi for the
correspondence and  Zhenya Skvortsov for pointing out a misleading
notation in Section 4.2 of the original version of the paper. V.D.
thanks Sasha Zhiboedov and Yi Pang for fruitful discussions.

\end{document}